# Ultrafast Optical Control of Multi-Valley States in 2D SnS

*Arqum Hashmi[1]\*, M. Umar Farooq[2], Mizuki Tani[3], Kazuhiro Yabana[4], Tomohito Otobe[3]† and Kenichi L. Ishikawa[1]†*

[1] *Department of Nuclear Engineering and Management, Graduate School of Engineering, The University of Tokyo, 7-3-1 Hongo, Bunkyo-ku, Tokyo 113-8656, Japan*

[2] *Department of Physics, Southern University of Science and Technology, Shenzhen, Guangdong*

[3] *Kansai Institute for Photon Science, National Institutes for Quantum Science and Technology (QST), Kyoto 619-0215, Japan*

[4] *Center for Computational Sciences, University of Tsukuba, Tsukuba 305-8577, Japan*

**Corresponding authors:**
\*hashmi@g.ecc.u-tokyo.ac.jp
†ishiken@n.t.u-tokyo.ac.jp, †otobe.tomohito@qst.go.jp

## ABSTRACT

We theoretically study the ultrafast optical control of multiple valley states in two-dimensional (2D) tin sulfide (SnS) monolayers, a member of the layered group-IV monochalcogenides, which is a promising class of materials for overcoming current challenges in valleytronics. By combining time-dependent density functional theory with Maxwell's equations, we simulate how both linearly and circularly polarized ultrashort laser pulses affect the electronic excitation dynamics and valley polarization in SnS. Our results reveal that the corrugated phosphorene-like crystal structure of SnS monolayers leads to the emergence of both linear and circular dichroism, allowing flexible manipulation of multi-valley excitation by simply adjusting the light polarization. Moreover, the interplay between broken inversion symmetry and spin-orbit coupling gives rise to distinct Berry curvature effects and spin-valley coupling, thereby enabling circular dichroism. Furthermore, we propose that tuning the carrier-envelope phase of few-cycle femtosecond laser pulses can achieve sub-cycle, ultrafast switching among multiple valley states. These findings not only deepen our understanding of valley dynamics in 2D materials but may also open new avenues for the development of valleytronic devices.

# I. INTRODUCTION

The rise of two-dimensional (2D) materials, such as Graphene [1], transition metal dichalcogenides (TMDCs) [2], and h-BN [3] has opened new pathways for studying valleytronics phenomena [4,5]. These materials feature time-reversal-connected valleys in 2D hexagonal lattices, where two energy-degenerate but inequivalent K and K′ points are distinguished by opposite spin angular momenta and Berry curvature [6]. To achieve valley polarization, which is crucial for valleytronics applications, time-reversal symmetry must be broken [7,8]. This can be achieved either by applying a magnetic field [5,9] or through dynamical processes like optical pumping with circularly polarized light [10,11]. In this context, valley-selective optical circular dichroism plays a pivotal role, as it enables the observation of the valley Hall effect, thus providing a pathway for the development of future valleytronics devices [12]. However, realizing practical applications faces challenges due to stringent experimental requirements such as cryogenic temperatures and strong magnetic fields [9,11,13].

Recent studies highlight bulk group-IV monochalcogenides (MC), MX (where M = Sn, Ge and X = S, Se), as a promising class of materials for addressing the present challenges in valleytronics [14–19]. Bulk MCs have layered structures, where the layers are weakly bonded through van der Waals interactions. They feature $D_{2h}^{16}$ space group with puckered layers akin to black phosphorus and exhibit high anisotropy between the armchair and zigzag directions [20,21,19]. Experimental findings have demonstrated valley-selective linear dichroism, revealing unique valley-specific optical transition rules [16,17]. Valleys along the Γ-X ($V_X$) and Γ-Y ($V_Y$) directions in reciprocal space can be selectively excited by *x*- and *y*-polarized light, respectively [14,16,17].

Interestingly, 2D layers of MC, consisting of mono- to a few layers, have also been synthesized experimentally [19,22,23]. It is theoretically predicted [20,24] that linearly polarized light can selectively excite paired valleys in MC monolayers, similar to their bulk counterparts. While previous studies have mainly focused on valley optical selection within the linear response [21,24–26], key questions remain unanswered regarding the nonlinear optical response. Specifically, will linear dichroism survive under strong laser fields, and what are the excitation dynamics? To fill these gaps, we will investigate how valley polarization evolves as the system transitions from linear to nonlinear optical regimes and how robust is linear dichroism under the strong field. Another significant aspect that has yet to be fully explored is the effect of transition from bulk to monolayer structures. This transition results in the loss of inversion symmetry, with monolayers belonging to the non-centrosymmetric $C_{2v}^7$ space group. In this symmetry-broken environment, spin-orbit coupling (SOC) generates an effective magnetic field that lifts the spin degeneracy. This raises the intriguing question: can monolayers of MCs exhibit spin(s)-valley($\tau$) coupling (s·$\tau$) physics, similar to their TMDC counterparts, where the helicity of light controls both spin and valley characteristics, ultimately giving rise to circular dichroism?

Here we focus on monolayer SnS, where the presence of heavy elements can result in significant SOC, making it an excellent candidate to study the valleytronics and spintronics applications. Using the time-dependent density functional theory (TDDFT), we aim to explore the potential for SnS monolayer to exhibit multi-valley states and identify the underlying mechanisms that govern such behavior. The outline of this paper is as follows. First, we revisit the laser polarization-dependent valley polarization within the linear regime. We then explore the degree of inter/intra valley polarization and robustness of the optical selection rule for the $V_X$ and $V_Y$ valleys in the nonlinear regime under the strong field. Next, we uncover a novel aspect of s·$\tau$ coupling in monolayer SnS, driven by Berry curvature and the presence of time-reversal spin splitting. Lastly, we demonstrate the ultrafast optical control of multi-valley selection on femtosecond timescales. Our study reveals multi-valley states and the coexistence of linear and circular dichroism in monolayer SnS, paving the way for advancements in valleytronics. Notably, the ultrafast valley polarization offers precise sub-cycle control of electron dynamics.

## II. THEORETICAL FORMALISM

We consider a monolayer SnS, located in the *xy* plane [Fig. 1(a)] at *z* = 0, subject to a normally incident light pulse, and model their interaction by combining TDDFT and Maxwell's equations, referred to as the 2D Maxwell-TDDFT approach [27], using the open source package SALMON (Scalable Ab-initio Light-Matter simulator for Optics and Nanoscience) [28,29]. The electronic response in SnS is calculated based on the TDDFT. The propagation of the incident light to the + *z* direction, treated as macroscopic electromagnetic fields, is described by Maxwell's equations in terms of the vector potential $\mathbf{A}(z,t)$ with the *x* and *y* component,

$$\frac{1}{c^2}\frac{\partial^2 \mathbf{A}(z,t)}{\partial t^2} - \frac{\partial^2 \mathbf{A}(z,t)}{\partial z^2} = \frac{4\pi}{c}\mathbf{J}(z,t) \qquad (1)$$

Here we use c.g.s. units. By assuming that that the monolayer is sufficiently thin, we can safely consider the vector potential is spatially uniform inside the thin layer. Thus, the electric current density $\mathbf{J}(z,t)$ in Eq. (1) can be approximated as

$$\mathbf{J}(z,t) \approx \delta(z)\mathbf{J}_{2D}(t), \qquad (2)$$

where $\delta(z)$ denotes the Dirac delta function, and $\mathbf{J}_{2D}(t)$ the 2D current density inside the thin layer. We treat it as a boundary value problem among the incident $\mathbf{A}^{(i)}$, reflected $\mathbf{A}^{(r)}$, and transmitted fields $\mathbf{A}^{(t)}$ as follows. At the monolayer position (*z* = 0), $\mathbf{A}(z,t)$ is continuous, and its derivative fulfills,

$$-\frac{\partial A(z,t)}{\partial z}\bigg|_{z=0_+} + \frac{\partial A(z,t)}{\partial z}\bigg|_{z=0_-} = \frac{4\pi}{c}J(z,t) \quad (3)$$

The continuity condition reads,

$$\mathbf{A}(z=0,t) = \mathbf{A}^{(t)}(t) = \mathbf{A}^{(i)}(t) + \mathbf{A}^{(r)}(t). \quad (4)$$

From Eqs. (3) and (4), we obtain the fundamental equation of the 2D approximation method,

$$\frac{d\mathbf{A}^{(t)}}{dt} = \frac{d\mathbf{A}^{(i)}}{dt} + 2\pi \mathbf{J}_{2D}[\mathbf{A}^{(t)}](t), \quad (5)$$

where $\mathbf{J}_{2D}[\mathbf{A}^{(t)}](t)$ stresses that $\mathbf{J}_{2D}(t)$ is determined by the vector potential at $z=0$ which is equal to $\mathbf{A}^{(t)}$.

The time evolution of electron Kohn-Sham orbitals in a unit cell driven by $\mathbf{A}^{(t)}$ is governed by the time-dependent Kohn-Sham (TDKS) equation. The TDKS for the Bloch orbital is expressed in the velocity gauge as,

$$i\hbar\frac{\partial}{\partial t}u_{b,\mathbf{k}}(\mathbf{r},t) = \left[\frac{1}{2m_e}\left\{-i\hbar\nabla + \hbar\mathbf{k} + \frac{e\mathbf{A}^{(t)}(t)}{c}\right\}^2 + \widehat{v_{ion}} + v_H(\mathbf{r},t) + v_{xc}(\mathbf{r},t)\right]u_{b,\mathbf{k}}(\mathbf{r},t), \quad (6)$$

where $u_{b,\mathbf{k}}(\mathbf{r},t)$ is the Bloch orbital with $b$ being the band index and $\mathbf{k}$ the 2D crystal momentum of the thin layer. The $\widehat{v_{ion}}, v_H, v_{xc}$ are ionic pseudopotential, Hartree potential and the exchange-correlation potential, respectively. Spin-orbit coupling (SOC) is introduced through the $j$-dependent nonlocal potential [30]. The 2D current density in Eq. (4) is obtained from the Bloch orbitals $u_{b,\mathbf{k}}(\mathbf{r},t)$ as follows:

$$\mathbf{J}_{2D}(t) = -\frac{e}{m_e}\int dz \int_\Omega \frac{dxdy}{\Omega} \sum_{b\mathbf{k}} u^*_{b\mathbf{k}}(\mathbf{r},t)\left\{-i\hbar\nabla + \hbar\mathbf{k} + \frac{e\mathbf{A}^{(t)}(t)}{c}\right\} \times u_{b,\mathbf{k}}(\mathbf{r},t) + \mathbf{J}_{ps} \quad (7)$$

where $\Omega$ is the 2D unit-cell area of the 2D material and $\mathbf{J}_{ps}$ is a current contribution from the nonlocal pseudopotential.

The excited electron population is defined as

$$\rho_{\mathbf{k}}(t) = \sum_{c,v}\left|\int_\Omega d^3\mathbf{r}\, u^{*}_{v,\mathbf{k}} u^{GS}_{c,\mathbf{k}+\frac{e}{\hbar c}\mathbf{A}^{(t)}(t)}\right|^2 \quad (8)$$

where $v$ and $c$ are the indices for the valence and conduction bands, respectively, and $u^{GS}_{b,\mathbf{k}}(\mathbf{r}) = u_{b,\mathbf{k}}(\mathbf{r}, t=0)$ is the Bloch orbital in the ground state.

**Computational Details**

The lattice constants along the zigzag and armchair directions of single-layer SnS are 4.08 Å and 4.28 Å. A vacuum layer of thickness 17 Å are attached to both sides of the layer. For electron-ion interaction, we employ the norm-conserving pseudopotential [31] for Sn and S atoms. The adiabatic local density approximation with Perdew-Zunger functional [32] is used for the exchange and correlation. The spin-orbit coupling (SOC) is implemented in a noncollinear mode [33,34]. We consider the linearly and circularly polarized laser field vector potential of the following waveform:

$$\mathbf{A}^{(i)}(t) = -\frac{cE_{max}}{\omega} f(t) \left[ \cos\left\{\omega\left(t - \frac{T_p}{2}\right) + \varphi\right\}\hat{\mathbf{x}} + \varepsilon \sin\left\{\omega\left(t - \frac{T_p}{2}\right) + \varphi\right\}\hat{\mathbf{y}} \right], \quad (9)$$

where $\omega$ is the average frequency, $E_{max}$ is the maximum amplitude of the electric field, $\varphi$ is a carrier-envelope phase (CEP) and $T_P$ is the foot-to-foot pulse duration, corresponding to $0.26T_P$ full-width-at-half-maximum pulse duration. The $\varepsilon = 0, 1$ and $-1$ describe the linear, right and left polarization respectively.

We use a $\cos^4$-type envelope function $f(t)$ given by,

$$f(t) = \begin{cases} \cos^4\left(\pi \frac{t - T_P/2}{T_P}\right), & 0 \leq t \leq T_P \\ 0, & otherwise \end{cases}, \quad (10)$$

We set the CEP $\varphi = 0$ unless otherwise stated (until Fig. 4). The spatial grid size and the number of $k$ points in the 2D reciprocal space are optimized through examination of the convergence of the calculation results; the $r$-grid is set to 26×26×120, and the $k$-mesh 35×35 in the Brillouin zone (BZ). The time step size is set to $5 \times 10^{-4}$ femtosecond.

**Dependence of Valley Polarization on Laser Polarization**

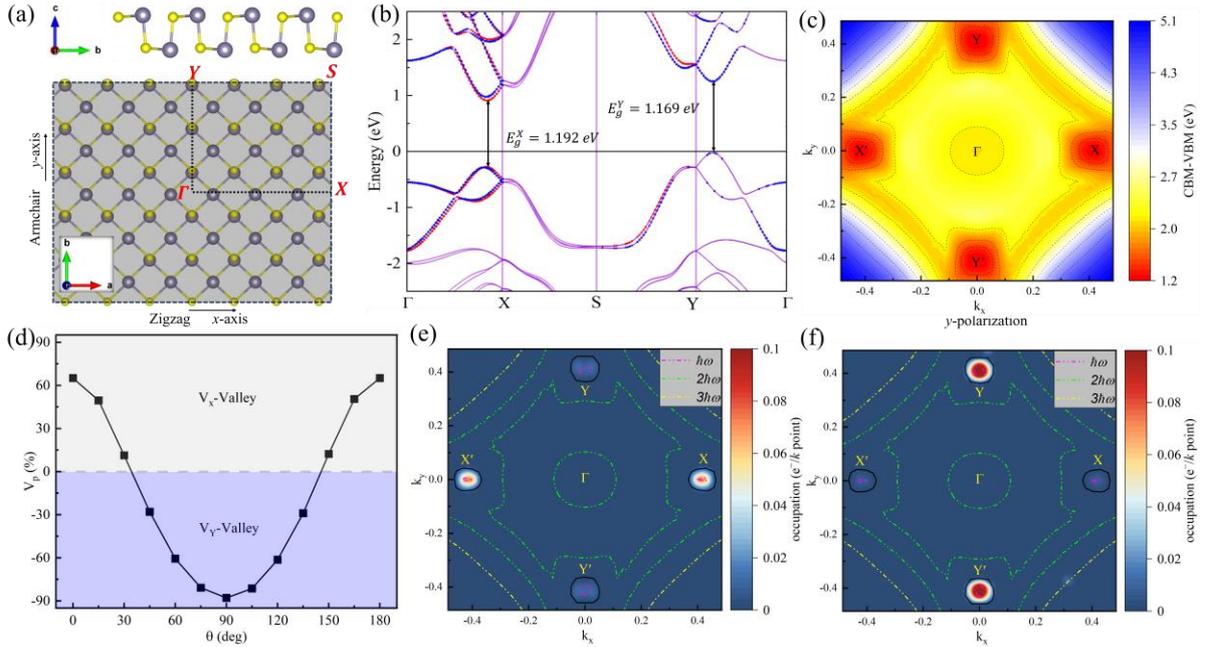

**Fig. 1:** (a) Schematic illustration of the SnS monolayer crystal structure. The upper panel shows the side view and the lower panel shows the top view, along with the respective BZ and high symmetry points. The grey (yellow) spheres represent Sn (S) atoms. (b) Electronic band structure with SOC along high symmetry points. The red and blue dots correspond to $S_z = \uparrow$ and $S_z = \downarrow$, respectively. (c) 2D map of the energy difference between the SnS CBM and VBM as a function of the wave vector $\mathbf{k}$ in the first BZ. (d) Polarization-dependent valley polarization. The positive values show the carrier density on the $V_X$-valley while negative values show that carrier density is concentrated on $V_Y$-valley. $\mathbf{k}$-resolved electron populations at the end of the pulse for (e) $\theta = 0°$ and (f) $\theta = 90°$ at weak intensity of $10^9$ W·cm$^{-2}$. The electron population is summed over the entire conduction band and the dashed lines indicate the resonances, where the energy difference between the valence and conduction band corresponds to the photon energy $\hbar\omega$, $2\hbar\omega$ or $3\hbar\omega$.

SnS monolayer is isoelectronic and has a structure similar to corrugated phosphorene as shown in Fig.1 (a). However, the presence of two different atomic species lowers crystal symmetry. The relative displacement between tin and chalcogenide atoms breaks the mirror symmetry along the armchair direction (*b*-axis) while preserving it along the zigzag direction (*a*-axis). Fig. 1(b) shows the electronic band structures at high-symmetry points, revealing the valence and conduction band exhibit two distinct maxima and minima along the Γ-X and Γ-Y sections of the BZ. Band profiles for the topmost valence (VBM) and the bottom conduction (CBM) bands were extracted from DFT, and the energy difference between CBM and VBM is illustrated in Fig. 1(c). These profiles indicate the positions of potential direct transitions for electrons from the valence to the conduction band at locations X(X') and Y(Y'), named $V_X$ and $V_Y$ valleys, respectively, with nearly identical band gaps whose values are indicated in Fig. 1(b). These results are consistent with prior studies [17,20,24,35].

Let us begin by analyzing how the laser-induced valley polarization changes as we vary the polarization direction of linearly polarized pulses within the *ab* plane, or *xy* plane. The valley polarization is defined as,

$$V_p = (\rho_X - \rho_Y)/(\rho_X + \rho_Y), \qquad (11)$$

where, $\rho_{X(Y)}$ represents the electron densities obtained by integrating the conduction band electron population around the *X(Y)* points inside the black dashed inner circle around *X(Y)* in Fig. 1(c). We use a photon energy of 1.10 eV, which is close to the bandgaps at the valleys and corresponds to a wavelength of 1127 nm. The foot-to-foot pulse length is set to 8 optical cycles, or 7.85 fs full-width-at-half-maximum (FWHM).

In this subsection, we consider the linear response, using a sufficiently weak intensity of $10^9$ W·cm$^{-2}$. Fig. 1(d) shows that the maximum valley polarization is achieved along the X-Γ-X' ($\theta = 0°$, *x*-polarization) and Y-Γ-Y' ($\theta = 90°$, *y*-polarization) directions, indicating strong valley selectivity in high-symmetry directions. This observation confirms the valley-related optical selection rule that the valley polarization in the momentum space is maximized when the excitation polarization matches the corresponding crystal orientation in the real space. The magnitude of the valley polarization is larger in the $V_Y$ valley ($\theta = 90°$) than in the $V_X$ valley ($\theta = 0°$). This difference can be attributed to the fact that an *x*-polarized light can populate both $V_X$ and $V_Y$ [Fig. 1(e)], though the $V_X$ valleys are excited more strongly. On the other hand, a *y*-polarized light primarily excites the $V_Y$ valleys and cannot excite the $V_X$ valley [Fig. 1(f)], since there is no coupling between the valence and conduction band at $V_X$ in the *y*-direction [24]. This observation is consistent with preceding works [16,17].

**Linear Dichroism Under Strong Field**

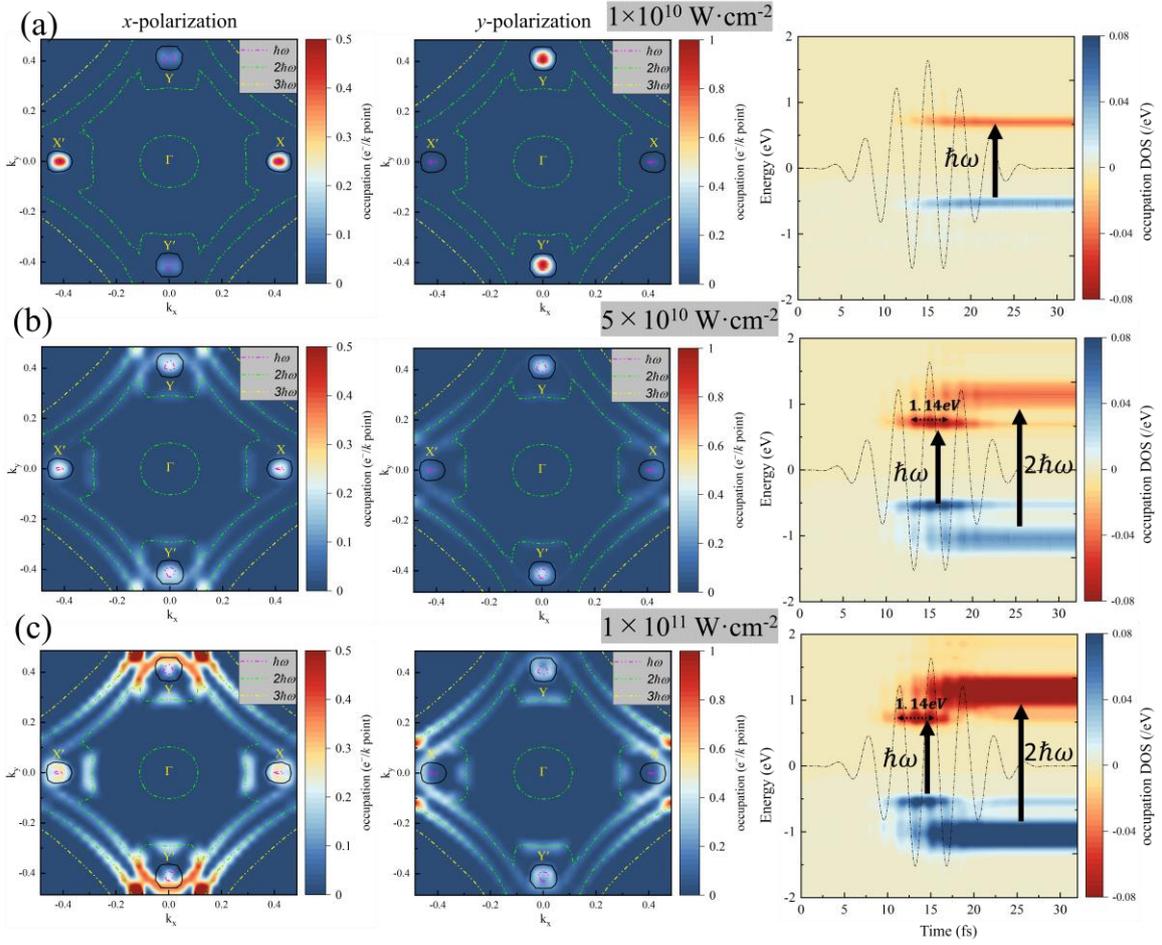

**Fig. 2:** Valley selective excitation by linearly polarized light in the first BZ at the end of the pulse for Γ-X (*x*-polarization) and Γ-Y (*y*-polarization) directions and the time-dependent evolution of the energy-resolved electron-hole distribution, projected onto ground state Kohn-Sham orbitals (far right panels) for (a) $1\times10^{10}$ W·cm$^{-2}$ and (b) $5\times10^{10}$ W·cm$^{-2}$ and (c) $1\times10^{11}$ W·cm$^{-2}$. The term occupation refers to the change in electron and hole occupation over time. The dashed line shows the applied electric field.

Let us next investigate how the valley polarization is affected if we increase the laser intensity to the nonlinear regime. To explore this, we apply laser pulses linearly polarized along the Γ-X (θ = 0°) and Γ-Y (θ = 90°) directions, with the same parameters as in the previous subsection, except for intensity. Fig. 2 shows the ***k***-resolved excited electron population distribution after the pulse, highlighting the valley selection $V_X$ and $V_Y$ through distinct electric field polarization (left and middle columns for *x* and *y*-polarization, respectively). The time evolution of the energy resolved electron-hole distribution for *y*-polarization is also presented in the right column. The behavior is similar for *x*-polarization. At an intensity of $10^{10}$ W·cm$^{-2}$, *x/y*-polarized light primarily excites the corresponding $V_X/V_Y$ valley [Fig. 2(a)]. The excited carriers are located precisely where the interband energy difference is $\hbar\omega$, indicating direct single-photon excitation from VBM to CBM at the X and Y points. The energy-resolved electron-hole distribution also shows the dominance of single-photon absorption over the entire pulse duration. As the intensity increases to $5\times10^{10}\,\text{W}\cdot\text{cm}^{-2}$ [Fig. 2(b)], *x/y*-polarized light excites not only the $V_X/V_Y$ valley but also the $V_Y/V_X$ valley, involving both single- and two-photon ionization. In addition, the regions between the valleys are slightly excited along the two-photon absorption lines, indicated by green dash-double-dotted lines. The single-photon absorption peak remains prominent up to ~ 20 fs and exhibits oscillations. These oscillations align with the field maxima and correspond to the applied field frequency, suggesting they may arise from virtual excitations. A subsequent decline in the peak is observed by the end of the pulse, attributed to Rabi oscillations. Our calculation of the ponderomotive energy (~11 meV) shows it is too small to close the one-photon channel, further supporting that the reduction is purely due to Rabi oscillations. At an even higher intensity of $10^{11}\,\text{W}\cdot\text{cm}^{-2}$, we see substantial excitation in broader regions, particularly, with increased excitation along the X-Y direction [Fig. 2(c)]. Remarkably, *x/y*-polarized light excites electron around the $V_Y/V_X$ valley more strongly than around the $V_X/V_Y$ valley, in striking contrast to the low intensity cases. Indeed, the momentum matrix elements for the VBM to CBM transition are around the perpendicular valleys than around the parallel valleys, although they have significantly lower occurrence rate for lower intensities since these correspond to allowed indirect transitions [35]. One can see that as we increase the field strength, one-photon absorption reduces progressively. In the high-intensity case considered here, two-photon absorption takes the lead in the electron-hole distribution over time, and higher-energy states become more important than VBM and CBM (not shown in the figure).

**Inter/Intra Valley Polarization & Electron Dynamics**

As we have seen above, *x/y* polarized light does not only excite the $V_X/V_Y$ valley but also the $V_Y/V_X$ valley, especially in the nonlinear regime. In order to analyze the relation between laser polarization and valley excitation in more detail, let us define inter-valley and intra-valley polarization.

Inter-valley polarization is defined as

$$P_{inter}^x = \frac{\rho_x^X - \rho_x^Y}{\rho_x^X + \rho_x^Y}, \qquad P_{inter}^y = \frac{\rho_y^Y - \rho_y^X}{\rho_y^Y + \rho_y^X}, \qquad (12)$$

where $\rho_i^V$ (where *i=x, y* and *V=X, Y*) represents the electron densities in the *V* valley under the *i*-polarized excitation pulse. $P_{inter}^x$ and $P_{inter}^y$ are the valley polarization [Eq. (11)] for the *x* and *y* polarized pulses, respectively. On the other hand, intra-valley is defined as

$$P_{intra}^X = \frac{\rho_x^X - \rho_y^X}{\rho_x^X + \rho_y^X}, \qquad P_{intra}^Y = \frac{\rho_y^Y - \rho_x^Y}{\rho_y^Y + \rho_x^Y} \qquad (13)$$

The inter-valley polarization measures the difference in excitation between the two valleys ($V_X$ and $V_Y$) for a given laser polarization, which corresponds to comparing experimental photoluminescence (PL) intensities between the valleys. The intra-valley polarization quantifies the effect of laser polarization direction on the excitation of each valley, corresponding to comparing PL intensities under parallel and cross-polarization. In other words, the former evaluates the anisotropy between $V_X$ and $V_Y$, while the latter determines the emission characteristics and anisotropy within each valley.

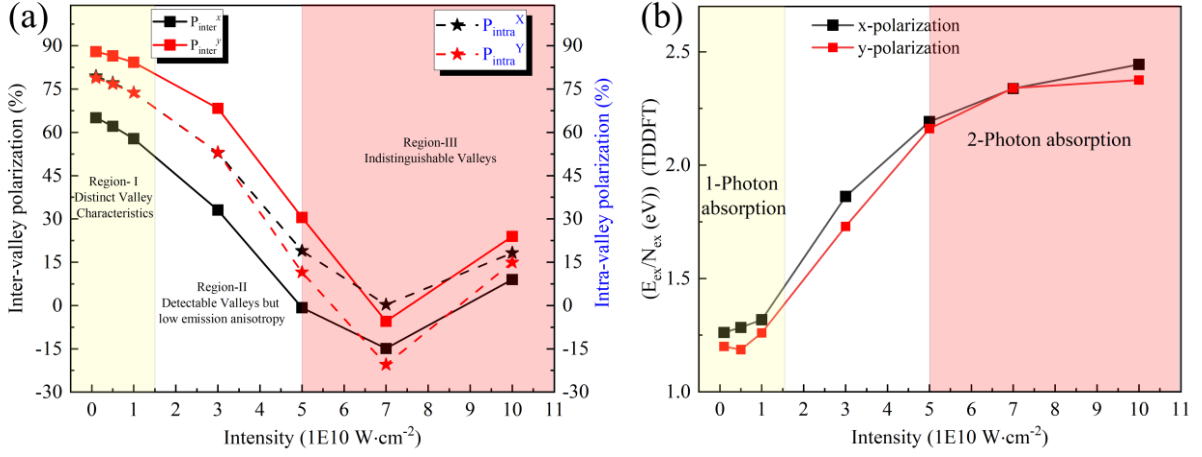

**Fig. 3:** (a) Inter- and intra-valley polarization vs. laser intensity, illustrating three regions from strong valley polarization to suppressed valley polarization. (b) Change in photon excitation energy ($E_{ex}/N_{ex}$).

The analysis of inter-valley and intra-valley polarizations reveals that $P_{inter}^y$ is consistently higher than $P_{inter}^x$ over the entire intensity range examined, while $P_{intra}^X$ and $P_{inter}^y$ have similar values. In addition, we can identify three distinct intensity regions [Fig. 3(a)] as follows:

**Region I**: At weak field strengths ($1\times10^9$- $1\times10^{10}$ W·cm$^{-2}$), both inter- and intra-valley polarization exhibit high values, indicating strong valley selectivity and efficient linear dichroism, though they slightly decrease with increasing intensity.

**Region II**: At intermediate field strengths ($3\times10^{10}$- $5\times10^{10}$ W·cm$^{-2}$), inter-valley polarization decreases significantly, and linear dichroism weakens, as intensity increases. In this intensity region, this occurs as the pulse excites both valleys comparably, reducing asymmetry. The inter-valley polarization at the $V_X$ valley almost vanishes at $5\times10^{10}$ W·cm$^{-2}$, while it remains positive (~30 %) at $V_Y$; there is still a noticeable difference between the $V_X$ and $V_Y$ valleys.

**Region III**: At strong field strengths ($> 5\times10^{10}$ W·cm$^{-2}$), the valley optical selection rules break down. Both the inter- and intra-valley polarization become even negative at $7\times10^{10}$ W·cm$^{-2}$, which indicates that the $V_Y$ and $V_X$ valleys are excited more by orthogonal polarization than by parallel polarization. Interestingly, valley polarization re-increases and becomes positive again with further increase in intensity. However, as we have seen in Fig. 2(c), the excitation areas around the $V_X$ and $V_Y$ valleys expand and are no longer completely isolated rendering valleys indistinguishable.

To further elucidate the reduction in valley polarization and linear dichroism, we plot the absorbed energy, evaluated as work done by laser field, per excited electron vs. laser intensity in Fig. 3(b). At low field strength (Region-I), the absorbed energy per electron is approximately 1.2~1.3 eV, close to the photon energy (1.10 eV), signifying that single-photon absorption dominates the photoexcitation process. Single-photon absorption highlights the optical transitions expected at the direct bandgaps located at the X and Y points of the BZ, preserving linear dichroism, since the transition probabilities are directly related to the polarization of the light. The absorbed energy per electron increases in the intermediate region (Region-II), which suggests the onset of two-photon absorption, leading to a reduction of valley polarization. At higher field strengths (Region-III), two-photon absorption is the main excitation path, as we have seen in Fig. 2. This detailed analysis shows that the key factor driving the change in valley polarization is the transition from single-photon to two-photon absorption. Two-photon absorption follows selection rules distinct from those for single-photon absorption, resulting in increased electron populations in the valleys on the transverse axis, thereby rendering valley selection irrelevant. Our findings suggest that linear dichroism only prevails under linear response and is suppressed by two-photon excitation.

**Berry Curvature and Spin-Valley Coupling**

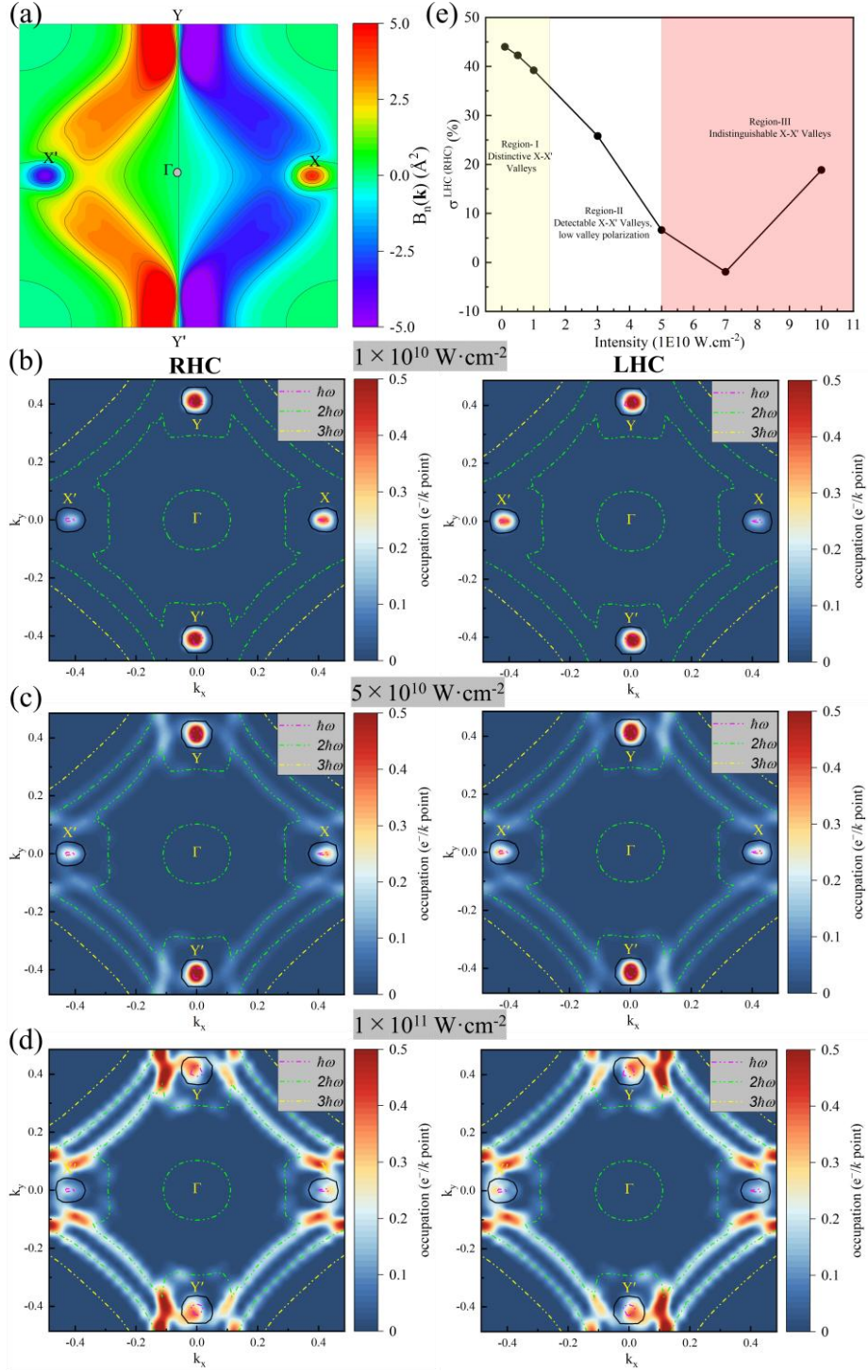

**Fig. 4:** (a) Momentum resolved Berry curvature (calculated by OpenMX [36]), illustrating positive and negative peaks at X and X' points. Distribution of k-resolved electron populations by circularly polarized light at intensity (b) $1\times10^{10}$ W·cm$^{-2}$, (c) $5\times10^{10}$ W·cm$^{-2}$ and (d) $1\times10^{11}$ W·cm$^{-2}$. The left and right panels show excited carrier population under right and left-handed circular polarization. (e) Circular valley polarization control relative to field strength.

Berry curvature [37–40] is an essential physical property in valleytronics, capturing the geometric aspects of wavefunctions and acting like a magnetic field in the momentum space. Berry curvature is given by $B_n(\boldsymbol{k}) =$ $\nabla \times W_{n\mathbf{k}}$, where $W_{n\mathbf{k}} = -i\langle u_{n\mathbf{k}}|\frac{\partial}{\partial \mathbf{k}}|u_{n\mathbf{k}}\rangle$ and $|u_{n\mathbf{k}}\rangle$ is a normalized Kohn-Sham orbital, and $n$ the band index. Figure 4(a) shows the $k$-resolved Berry curvature $B_n(\mathbf{k})$ integrated over the whole valence bands across the 2D BZ. The non-vanishing $B_n(\mathbf{k})$ in the SnS monolayer arises from its non-centrosymmetric lattice structure. The Berry curvature properties are directly tied to the symmetry of the orthorhombic SnS monolayer. Mirror symmetry is broken along the y-direction (armchair), while it is preserved along the x-direction (zigzag) [Fig. 1(a)]. As a result, $B_n(\mathbf{k})$ behaves as an odd function of $k_x$ and an even function of $k_y$. Therefore, the Berry curvature is positive and negative at X and X' valley-points, respectively. In addition, the lack of inversion symmetry leads to spin splitting of 6 and 70 meV on VBM and CBM, respectively, [Fig. 1(b)] with opposite spin at the time-reversal points X and X' owing to SOC. In contrast, the Y and Y' points have vanishing Berry curvature and spin splitting, possibly due to the absence of mirror symmetry with respect to the *x* axis. The Zeeman-type spin splitting leads to the $V_X$ valley splitting into the two degenerates but inequivalent valleys at X and X' points analogous to the K and K' valleys in TMDCs. As a result, the SnS monolayer exhibits s · τ coupling with optical selectivity determined by the spin index [41,42], hence valley selectivity can be achieved by the helicity of light.

**Circular Dichroism in 2D SnS**

Let us consider circularly polarized pulses with the same wavelength and pulse duration as those employed for linearly polarized pulses. As mentioned above, the combination of broken inversion symmetry and time-reversal symmetry ensures that the energy of the spin-down state at X' is equal to the energy of the spin-up state at X, i.e., E↓(X') = E↑(X), effectively linking spin-down and spin-up states across the X' and X points. Due to this selection rule, interband transitions at the X and X′ valleys can be selectively excited by right-hand circularly (RHC) and left-hand circularly (LHC) polarized light, respectively, yielding equivalent magnitudes of circular valley polarization [Eq. (15)], albeit in opposite valleys. Figure 4(b - d) illustrates the excited carrier population under the RHC and LHC light. When RHC is applied, mainly spin-up electrons are excited to the conduction band at X (left panel), while mainly spin-down electrons are excited at X' when LHC (right panel) is applied. This results in a robust circular dichroic effect in the excited state populations. The circular dichroism is evaluated in terms of circular valley polarization, which is defined similarly to that in TMDCs as follows:

$$\sigma^{RHC(LHC)} = \frac{\rho_X - \rho_{X'}}{\rho_X + \rho_{X'}} \qquad (15)$$

where $\rho_X$ and $\rho_{X'}$ are electron densities around the X and X' valley points, respectively. Note that, unlike TMDC where in an ideal case the $\sigma^{LHC}/\sigma^{RHC}$ gives ~100% circular dichroism [10,43], the SnS monolayer exhibits strong but not absolute circular dichroism. However, the Berry curvature in the SnS monolayer allows the electric detection of $s \cdot \tau$ polarization through an anomalous velocity perpendicular to the applied electric field ($\sim \mathbf{E} \times B_n(\mathbf{k})$). Charge carriers with $s \cdot \tau = \pm 1/2$, excitable by left- and right-circularly polarized light, respectively, will exhibit opposite transverse conductivity ($\sigma_{xy}$), which can be detected via the valley Hall effect similar to TMDC. We observe a maximum circular valley polarization of 44% in the weak field limit (Fig. 4(e)). The relatively moderate circular valley polarization compared to the linear polarization case [Fig. 3(a)] can be attributed to the small spin splitting of VBM and CBM at the X(X') valleys, which limits the degree of polarization. Figure 4(e) shows a trend that is strikingly similar to the trend we have seen for linear polarization [Fig. 3(a)]; Under weak field where the X and X' valleys are distinct, circular dichroism initially appears but decreases with increasing field intensity. At resonant frequencies, both linear and circular polarizations exhibit single-photon transitions, localizing carriers at the valley points. However, at intensities $\geq 5 \times 10^{10}$ W·cm$^{-2}$, two-photon absorption becomes prominent, causing circular valley polarization to decrease as the population shifts toward regions adjacent to the valleys. At the highest field intensity region, the nonlinear interaction further spreads excited electrons beyond the X and X' points, similar to the behavior observed for linear polarization. The SnS monolayer has not only $V_X$ and $V_Y$ valleys but also X-X' valley states, each governed by distinct optical selection rules. This allows the formation of multi-valley states, providing additional valley degrees of freedom within a single system. Moreover, these valleys can be selectively accessed via linear and circular dichroism, providing a way to control valley-specific excitations.

**Ultrafast Optical Control of Multi-Valley States**

The rectangular unit cell, as marked by black squares in Fig. 5(a)-(c), positions the X-X' (Y-Y') valleys near the BZ boundaries. As a result, the X-X' (Y-Y') distance on the $k$ plane across the BZ boundaries (0.16 a.u.) is much smaller than it appears within the BZ (0.66 a.u.). This may result in faster inter-valley scattering than in TMD. Inspired by this conjecture, let us now explore the possibility of controlling carrier excitation at multiple valley points through the polarization and carrier-envelope phase $\varphi$ [see Eq. (9)] of linearly polarized ultrashort laser pulses. Fig. 5 presents the dependence of valley polarization on the laser polarization angle θ for $\varphi = 0°$ and $\varphi = 90°$. Fig. 5(a-c) displays the $k$-resolved electron population at the end of the pulse with 0.4 eV photon energy (much smaller than the bandgap) and $5 \times 10^{10}$ W·cm$^{-2}$ peak intensity. Practically no valley asymmetry is observed at $10^{10}$ W·cm$^{-2}$ (not shown). At $5 \times 10^{10}$ W·cm$^{-2}$, for which the vector potential $A(t)$ (0.16 a.u.) is comparable to the distance between X-X' (or Y-Y') separation across the BZ boundaries, a clear valley asymmetry is observed. When the laser polarization is parallel to Γ-X (θ = 0°) and $\varphi = 0°$, both X and X' valleys are equally excited [Fig. 5(a), upper panel]. In contrast, for θ = 0° and $\varphi = 90°$, a strong valley asymmetry is observed [Fig. 5(a), lower panel]. Consequently, we see a substantial valley polarization [Fig. 5(d)]. When the laser is polarized parallel to Γ-Y direction (θ = 90°) and φ = 90° [Fig. 5(b)], either Y or Y' is more excited than the other. Thus, $V_Y$ valley results in two inequivalent valleys, and, therefore, in spite of the absence of opposite spin splitting and Berry curvature, 15% valley polarization is achieved at Y-Y' points. The electron distribution for θ = 180° [Fig. 5(c)] is reversed with respect to the Y-Y' line, compared to the case of θ = 0°, as is expected from the symmetry of the band structure, and the opposite valley (see Fig. 5(c)) is selected. For this non-resonant $\hbar\omega \ll E_g$ (bandgap) process, the absorbed energy per electron is ~1.7 eV, corresponding to ~ 4 photons. Thus, while the valley selection rule for the case of circularly polarized light with a photon energy close to the bandgap depends on Berry curvature as we have seen in the previous subsection, valley polarization under a linearly polarized ultrashort laser pulse with a photon energy much smaller than the bandgap is driven by multiphoton excitation, Bragg scattering, and the interference of the wave packets [44–46].

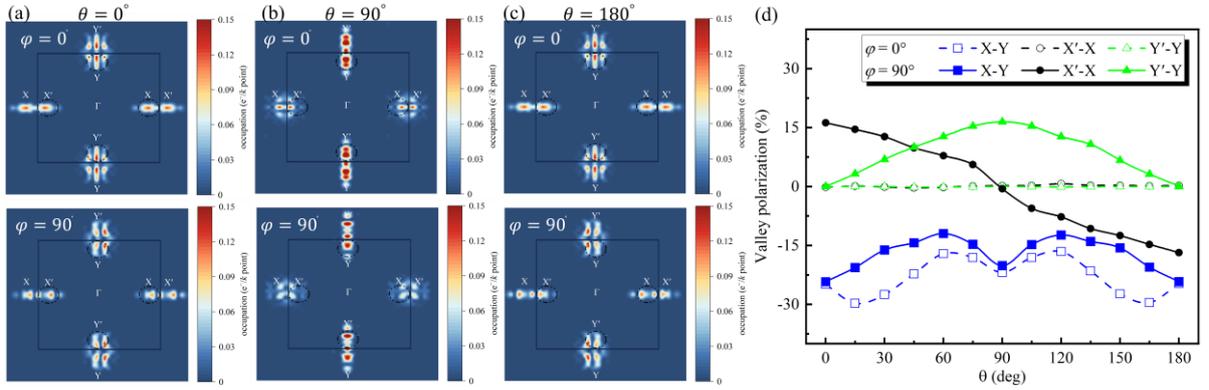

**Fig. 5:** Valley selection by ultrashort linearly polarized 3-cycle pulses with a photon energy of 0.4 eV and field strength of $5\times10^{10}$ W·cm$^{-2}$. Distribution of k-resolved electron populations at the end of the pulse for polarization direction (a) θ = 0° (x-polarized), (b) and θ = 90° (y-polarized) and (c) θ = 180° (-x-polarized). The upper panels are for CEP = 0° while the lower panels are for CEP = 90°. (d) Valley polarization as a function of θ for X-Y, X-X', and Y-Y' valleys for CEP = 0° and 90°. To calculate the valley polarization, the electron densities are obtained by integrating the conduction band electron population around the X(Y) points within black dashed circles.

Figure 5(d) shows the valley polarization as a function of in-plane laser polarization angle θ. The valley polarization is calculated by Eq. (11), in which the electron densities are obtained by integrating the conduction band electron population around valley points (X-Y, X-X', and Y-Y') as indicated by black dashed circles in Fig. 5 (a)-(c). As mentioned above, the valley polarization for X-X' (Y-Y') vanishes in the weak intensity limit ($10^{10}$ W·cm$^{-2}$). On the other hand, we see from Fig. 5(d) that as the intensity increases to $5\times10^{10}$ W·cm$^{-2}$ the nonlinear processes lead to a substantial valley polarization. This behavior contrasts with the previously described linear and circular dichroism, which diminishes with increasing intensity. For CEP = 90°, the valley polarization along the X-X' direction is maximized at θ = 0° (∥ x) and monotonically decreases as θ increases, eventually reaching zero at θ = 90° (∥ y); the valley polarization along the Y-Y' direction is maximized at θ = 90° (∥ y) and monotonically decreases as θ moves away from 90°, eventually vanishing at θ = 0° and 180° (∥ x). Thus, the valley polarization is maximized (vanishes) when the laser polarization is parallel (perpendicular) to the valley direction. The maximum valley polarization between X-X'(Y-Y') is 16%, which is expected to be observed experimentally [47,48]. Though less prominent than in the case of the resonant 1.1 eV photon energy (Fig. 2), we still obtain a sizable inter-valley (X-Y) polarization of up to ~30% [Fig. 5(d)]. Its dependence on $\varphi$ and $\theta$ is more complicated than for the X-X'(Y-Y') polarization. Hence, these findings indicate that tailoring precisely the wave form of ultrashort laser pulses enables unprecedented control over the electron dynamics across multiple valleys on a timescale even faster than electron-electron and electron-phonon scattering.

## IV. CONCLUSION

Using TDDFT calculations, we have investigated valley control in 2D SnS monolayers. In contrast to conventional materials that typically rely on a single valley pair, SnS offers the possibility of exploiting multiple valley states, including X-X', Y-Y', and X-Y, thereby significantly enhancing the degrees of freedom for valley manipulation. Moreover, SnS monolayers exhibit both linear and circular dichroism due to the corrugated phosphorene-like crystal structure, enabling flexible valley-selective excitation by simply adjusting the light polarization. Furthermore, the interplay between inversion symmetry breaking and strong spin-orbit coupling in SnS produces Berry curvatures of opposite signs at the X and X' points. As a result, circularly polarized light can induce spin-valley coupling, allowing for the simultaneous manipulation of both spin and valley. These features are unique in 2D SnS. We have also proposed a method of rapidly switching among multiple valley states on a sub-cycle timescale using femtosecond laser pulses. Precise CEP manipulation enables the selective excitation of multiple valley states, opening new possibilities for ultrafast valley-based technologies.

**Notes**

The authors declare no competing financial interest.

**ACKNOWLEDGMENT**

This research is supported by JSPS KAKENHI Grant No. JP20H05670. This research is also partially supported by MEXT Quantum Leap Flagship Program (MEXT Q-LEAP) under Grant No. JPMXS0118067246 and JPMXS0118068681, and JSPS KAKENHI Grant Nos. JP20H02649, JP22K13991, JP22K18982, JP24H00427,

and JP24K01224, and JST-CREST under Grant No. JPMJCR16N5. The numerical calculations are carried out using the computer facilities of the Fugaku through the HPCI System Research Project (Project ID: hp220120, hp240124), SGI8600 at Japan Atomic Energy Agency (JAEA), and Wisteria at the University of Tokyo under Multidisciplinary Cooperative Research Program in CCS, University of Tsukuba and JSPS KAKENHI Grant, JP22K13991.